\documentclass[anonymous,runningheads]{llncs}
\usepackage[T1]{fontenc}
\usepackage{hyperref}
\usepackage{booktabs}
\usepackage{tabularx}
\usepackage{tablefootnote}
\usepackage{comment}
\usepackage{fancyvrb}
\usepackage{subcaption}
\usepackage{listings}
\usepackage{makecell}
\usepackage{amsmath}
\usepackage{csquotes}
\usepackage{xurl}
\lstdefinelanguage{PHP}{
  morekeywords=[1]{
    abstract, and, array, as, break, callable, case, catch, class, clone,
    const, continue, declare, default, do, echo, else, elseif, empty,
    enddeclare, endfor, endforeach, endif, endswitch, endwhile, enum,
    eval, exit, extends, final, finally, fn, for, foreach, function,
    global, goto, if, implements, include, include_once, instanceof,
    insteadof, interface, isset, list, match, namespace, new, or, print,
    private, protected, public, readonly, require, require_once, return,
    static, switch, throw, trait, try, unset, use, var, while, xor, yield,
    yield from
  },
  morekeywords=[2]{
    int, float, string, bool, void, null, mixed, never, object, self,
    parent, true, false, iterable, array
  },
  morekeywords=[3]{
    __CLASS__, __DIR__, __FILE__, __FUNCTION__, __LINE__, __METHOD__,
    __NAMESPACE__, __TRAIT__
  },
  sensitive=true,
  morecomment=[l]{//},
  morecomment=[l]{\#},
  morecomment=[s]{/*}{*/},
  morecomment=[s]{/**}{*/},
  morestring=[b]",
  morestring=[b]',
  morestring=[s]{<<<}{;},
}
\usepackage[ruled,vlined]{algorithm2e}
\usepackage{xcolor}

\usepackage{cleveref}

\crefname{lstlisting}{listing}{listings}
\Crefname{lstlisting}{Listing}{Listings}
\crefname{algorithm}{algorithm}{algorithms}
\Crefname{algorithm}{Algorithm}{Algorithms}
\crefname{section}{section}{sections}
\Crefname{section}{Section}{Sections}
\crefname{appendix}{appendix}{appendices}
\Crefname{appendix}{Appendix}{Appendices}

\usepackage{graphicx}
\usepackage{color}

\newif\ifanon
\anonfalse 

\begin{document}


\title{Poster: To Play or Not to Play: \\ Insights and Lessons Learned from \\ 20~Years of CTFs with ENOFLAG}
\titlerunning{Poster: Insights and Lessons Learned from 20 Years of CTFs}
%
\author{
Jörg Schneider\inst{1}\orcidID{0009-0006-3115-1946} \and     
Sebastian Neef\inst{1}\orcidID{0000-0003-3055-0823} \and
Sebastian Koch\inst{1}\orcidID{0009-0002-6770-1656}
}
\authorrunning{J. Schneider et al.}
%
\institute{Technische Universität Berlin, Einsteinufer 17, 10587 Berlin, Germany
}
\maketitle              

\newcommand{\ENOFLAG}{\emph{ENOFLAG}}
\newcommand{\ENOWARS}{\emph{ENOWARS}}
\newcommand{\BAMBICTF}{\emph{BAMBICTF}}

\begin{abstract}
Security contests in the form of CTF (Capture The Flag) exercises are nowadays a common way to learn cyber security. 20 years ago at DIMVA 2006 the on-site CTF \emph{CIPHER II} was one of the conference highlights and led to the foundation of the team \ENOFLAG. In this poster, we reflect on the changes in the CTF gameplay and report on lessons learned while running an academic CTF team for 20 years.

\end{abstract}

\section{Introduction}
Capture The Flag (CTF) is commonly known as a children's game to foster important skills.
Gamification can serve as a useful tool to convey students new concepts and knowledge in a playful way and has therefore been used for IT security classes \cite{McDaniel16}.
For more than 20 years, the IT security community has adopted CTF-style competitions to share and teach IT-security knowledge in academic and non-academic settings, with almost a thousand CTF teams competing against each other in several hundred competitions per year\footnote{\url{https://ctftime.org/}}.

The CTF team \ENOFLAG{} was founded as an academic team in 2006 
at Technical University of Berlin (Germany), to participate in \emph{CIPHER II} at DIMVA 2006 \cite{cipher_dimva2006_report_2006}. 
Over the past 20 years, this team organized dozens of attack-defense and jeopardy-style CTF competitions, and participated in countless on-line or on-site CTF qualifiers and finals all over the world, ranking among the top 5 teams in its country.\footnote{CTFtime: Team ENOFLAG \url{https://ctftime.org/team/1438}} 

Over the 20 years, we witnessed shifts in how CTFs are organized and played, including modern challenges, such as AI usage.
Thus, this poster aims to inform and extend the body of knowledge as follows:
\begin{itemize}
    \item Insights into the evolution of CTFs over 20 years
    \item Lessons learned from organizing and partaking in CTFs as \ENOFLAG{} for 20 years
\end{itemize}

\section{Background}\label{sec:background}

\subsection{Capture The Flag Competitions}
\paragraph{Overview}
In the IT-security context, Capture The Flag (CTF) competitions are a contest where the organizers, which are often other CTF teams, prepare a set of intentionally vulnerable programs or systems (so-called \emph{challenges}) of varying difficulty. 
The participating CTF teams then study the challenges, identify the intentional vulnerabilities to obtain a hidden piece of information (the flag) after successful exploitation thereof. 
Successful retrieval of a flag and submission to the organizer will earn a team points on a scoreboard. 
After the competition ends, the highest scoring team wins. 
Two distinct competition types are common within the community: \emph{Jeopardy} and \emph{Attack-Defense (A/D)}.

\paragraph{Jeopardy-style}
In this game mode, the organizers prepare a large set of challenges in different IT-security categories, e.g., web-security, cryptography, reverse engineering, binary-exploitation (pwn), forensics, or others (miscellaneous).
Each category consists of at least 5 challenges of increasing difficulty and all participating teams get the same set of challenges.
During the competition's runtime of commonly 24 to 48 hours, the goal is to solve as many of these challenges as fast as possible. 

\paragraph{Attack-Defense style}
This game mode differs in the complexity of organization and interactivity between the participating teams. 
Contrary to Jeopardy-style CTFs, the organizers prepare a virtual machine with a set of around 8 vulnerable software services.
Each team gets their own identical copy of the virtual machine, which it \emph{also} must protect. 
This means identified vulnerabilities need to be patched on a team's machine without breaking its functionality, while at the same time exploiting them on other teams' machines to obtain their flags.
New flags are placed by the organizers for all services of all teams in a round-based fashion. 
Thus, exploits must be automated to run against all other teams each round at least once. 
The runtime of such interactive CTFs is around 8 hours, but can be longer in some cases, e.g. for local on-site finals.

\paragraph{Teams}
Participating teams can be roughly categorized into school, academic, and (professional) community teams.
While academic and community teams compete against each other, there are specific CTFs tailored to be more beginner-friendly with simpler tasks for (high)school teams.
Academic teams consist of (graduate) students, PhD students, and have respective supervisors (e.g. professors or instructors), placing the focus on the educational purpose of participation, while other teams might have different motivation to participate, e.g. to win prizes.
For on-site competitions, the team size can be limited to 5-10 participants, while on-line CTFs naturally cannot effectively limit the team size, so some teams participate with up to 50 or more team members.

\subsection{History of ENOFLAG}\label{sec:enoflag}

A CTF was planned as one of the highlights of the DIMVA 2006 conference in Berlin: The second edition of the \emph{CIPHER} contest organized by Lexi Pimenidis.\cite{cipher_dimva2006_report_2006} The CTF should not only run as an online competition, but also feature local teams competing on-site. However, at this time only a few teams played CTFs and none was located in Berlin. Therefore, new teams had to be formed. At TU Berlin, we gathered interested students from two security-related lectures and started a meeting series to prepare for the contest. This series was continued after the contest and the \emph{AG Rechnersicherheit}\footnote{translates to: working group computer security} still meets at the same time - every Tuesday at 6 p.m.\footnote{\url{https://www.enoflag.de/}}

The team \ENOFLAG{} participated in each of the following \emph{CIPHER} contests during DIMVA conferences until the series ended. The first contest won was the prestigious UCSB iCTF 2008\cite{Childers2010}.
However, the goal of the \ENOFLAG{} team was not to form a team of few elite players, but to be a broad inclusive team also for first time players. 

Up until today, \ENOFLAG{} has participated in over 200 CTFs, both A/D and Jeopardy, and has organized more than 20 competitions for public and non-public events\footnote{CTFtime: Team ENOFLAG \url{https://ctftime.org/team/1438}}. 
Their \ENOWARS{} A/D CTFs are the yearly result of the accompanying project course at the university, where students develop CTF challenges. 
At the semester's beginning, an introductory CTF (\BAMBICTF{}) with a handful of beginner-friendly challenges is organized by the instructors to convey the game mechanics and A/D concepts.

\section{Evolution of CTFs}\label{sec:ctf}
Over the past 20 years, \ENOFLAG{} regularly participated and organized CTF competitions and saw several changes in the evolution between 2006 and 2026.

\paragraph{Participation}
Back in 2006, the number of teams participating in CTFs was in the range of 20-25 teams. 
For example, the \emph{CIPHER II} CTF only had 19 teams, while the UCSB iCTF 2006, which was the largest A/D CTF at the time, only saw at most 25 participating teams. 
Vigna et al, who were responsible for the iCTF, described hosting more than 100 teams as a challenge \cite{3gse2014-10-years-of-ictf} in 2014.
However, nowadays, several hundred teams participating in a CTF is the new normal, even for the more resource-intensive A/D CTFs\footnote{\url{https://ctftime.org/}}. 
One reason is the steep increase in the number of active teams. For example, in 2026 there are now multiple active teams in the Berlin metropolitan area alone, while in 2006 a team had to be formed to participate in \emph{CIPHER II}. Additionally, the increase of participants is accompanied by advances in cloud computing, which allows the organizers and participating CTF teams to host their infrastructure in scalable cloud environments, removing local hardware or bandwidth limitations.   

\paragraph{A/D vs. Jeopardy}
We have also observed a change in the style of CTFs.
The amount of Jeopardy-style CTFs has surpassed the A/D-style today,  
which was dominant 20 years ago.
One reason for this could be that hosting Jeopardy-style CTFs is less complex and less resource-intensive than A/D, especially with community-made Jeopardy-style CTF platforms, such as CTFd\footnote{\label{ctfd}\url{https://github.com/CTFd/CTFd/}}.

\paragraph{Tooling for Participants}
With the advancement of computer hardware and cloud computing, the tooling has evolved as well. 
Today, A/D organizers are able to offer all teams the same hardware resources by spawning hundreds of cloud-hosted VMs, whereas in the past, each team had to host their VMs on their own limited hardware.
Furthermore, while \ENOFLAG{} used \emph{tcpdump} and \emph{wireshark} to analyze traffic or simple python scripts for exploitation in the past, full-scale traffic analysis and exploitation systems exist today (e.g. Arkime,\footnote{\url{https://github.com/ENOFLAG/EnoArkime}} 
OpenAttackDefenseTools\footnote{\url{https://github.com/OpenAttackDefenseTools}}).

\paragraph{Tooling for Organizers} From an organizer's perspective, the tooling also became easier. In the early days, not only the challenges but also the whole management framework and scoreboard was written from scratch most of the time. While this allowed unconventional scoring and gameplay mechanisms\cite{Childers2010}, it presented an entry barrier for smaller teams organizing their first CTF. In the last 10 years, multiple platforms for CTF game management emerged \cite{Kucek2020}.

\paragraph{Writeups}
Writeups are descriptions of how a CTF challenge was solved, which are shared with the CTF community.
They are a valuable resource to learn new techniques and the approaches taken by other CTF players.
While many of these writeups were published publicly in the past, e.g., on CTFtime\footnote{\url{https://ctftime.org/writeups}}
or GitHub\footnote{\url{https://github.com/ctfs}}, 
we have seen that writeups are published less frequently or shared within gated-communities (e.g., Discord-servers), making knowledge less accessible to the public.
For example, the last writeup-repository from the aforementioned GitHub community is from 2018, and less writeups are published on CTFtime after a CTF nowadays, e.g., for our recent ENOWARS CTFs\footnote{\url{https://ctftime.org/ctf/24}} 2019: 11, 2020: 0, 2021: 4, 2022: 0, 2023: 0, 2024: 0, 2025: 1. 

\paragraph{AI}
With recent developments in AI and its agentic capabilities, AI has had a transformative impact on CTFs.
Especially for Jeopardy-style CTFs, the AI agents can be provided the challenge description, files, and remote server to figure out and exploit the vulnerability. 
Krauq\footnote{\url{https://github.com/OpenAttackDefenseTools}}
and L.A.R.P\footnote{\url{https://ctftime.org/team/269755/}}
have demonstrated the feasibility of winning CTFs with agentic AI, solving challenges of varying difficulty and from all categories. 
Thus, CTF teams now either have to use AI as well, or their chances of winning become diminishingly small, as the speed and capabilities of AI outpace a single player.
The community is currently debating AI use for CTFs, but will eventually have to adapt to it, e.g. by asking for prompts that solve challenges instead of having participants solve them directly.

Additionally, CTFs had also an impact on the AI development. The traffic logs of CTFs were used as training data and benchmarks for intrusion detection systems. Especially, the DEFCON8 dataset is widely cited (e.g. \cite{Nehinbe2010}). On the other hand, CTF challenges are used to evaluate the performance of AI agents\footnote{\url{https://www.aisi.gov.uk/blog/our-evaluation-of-claude-mythos-previews-cyber-capabilities}}.

\section{Lessons Learned}\label{sec:lessonslearned}
\paragraph{20 years \ENOFLAG}
Our team has always aimed to keep an open culture for newcomers with the goal of learning \emph{something}, instead of becoming an elite CTF team with few top players focused on winning prizes.
For example, we have played many CTFs with more than 40 participants, many of which were beginners or from non-CS courses of study.
We have gone through several changes of generations in our team. 
Each semester, we have new students showing up to our meetings, or alumni joining us for popular CTFs. 

While the engagement varies over time, we think that our consistent weekly meetings, a mailing list and instant-messenger groups (e.g. on Signal) for communication, and occasional get-togethers are important pillars to keep a CTF team alive for many years.
Additionally, we recommend founding an official association, as the team went through several supporting professor changes, university building and room changes, and changes in active members.

\paragraph{Tooling}
Adopting new technology, such as cloud hosting, has drastically reduced the preparation efforts for our team. 
In the past, at least two people spent several days obtaining permissions from the university, configuring networks, preparing servers and services for one A/D CTF. 
Today, we can simply spawn our complete infrastructure in an automated way the day before. 
Similarly, the new powerful software stacks for traffic analysis and exploit execution have simplified our workflows during CTFs, which also benefits beginners by reducing the barrier to entry.

\paragraph{Teaching}
While there is an overlap between instructors for IT-security classes and the \ENOFLAG{} team, many team members also support teaching the classes in their spare time. 
For example, during the \ENOWARS{} projects, \ENOFLAG{} team members are helping out with the introductory \BAMBICTF{} and later play-test the \ENOWARS{} services during a test-run.
Thus, having a closely related CTF team affiliated with an IT-security chair can bring important benefits to both parties, e.g. finding candidates for teaching assistants or research positions.
Also students have reported CTF-style teaching to be \emph{fun} and \emph{effective} due to its hands-on problem-based learning approach. 

\subsection{Outlook}
CTF competitions will stay an interesting approach to conveying new concepts to students in an academic setting. 
However, with the rise of AI and its agentic abilities to autonomously solve CTF challenges, the style of CTFs or their prize pools will have to be re-evaluated.
We believe that the current goals of solving as many challenges as fast as possible lead to a race about having the most capable agentic system, which translates to spending the most money on AI subscriptions or dedicated hardware. 
Thus, the gap between well-funded professional and underfunded academic teams widens, motivating more AI use to solve challenges and thus drastically reducing the learning effects of manually trying to solve a CTF challenge.
Furthermore, such developments could lead to students losing interest in participating if the chances of winning against AI-supported teams become diminishingly small.
While AI use cannot be effectively banned and should rather be embraced as a new technology, the CTF community needs to find new game modes or approaches how to keep CTFs interesting and fun.

\section{Conclusion}\label{sec:conclusion}
Our poster shows the insights and lessons learned from organizing and participating in CTF competitions for 20 years. 
Technological changes will continue to shape how teams participate in CTFs and impose new challenges for the community to create fair, interesting and fun competitions for the best knowledge exchange.

\appendix

%
%
%
\bibliographystyle{splncs04}
\bibliography{bibliography}
\end{document}